\documentstyle[12pt]{article}
\begin{document}
\def\strut{\rule[-.5cm]{0cm}{1cm}}
\def\dspace{\baselineskip = .30in}

\title{\Large\bf Topologically Stable Z - Strings in the
Supersymmetric  Standard Model}

\author{{\bf G. Dvali}\\
Dipartimento di Fisica, Universita di Pisa and INFN,\\
Sezione di Pisa I-56100 Pisa, Italy\\
and\\
Institute of Physics, Georgian Academy of Sciences,\\
380077 Tbilisi, Georgia\\
E-mail: dvali@mvxpi1.difi.unipi.it\\
\and
{\bf Goran Senjanovi{\'c}}\\
International Centre for Theoretical Physics\\
Trieste, Italy\\
E-mail: goran@ictp.trieste.it}

\date{ }
\maketitle

\begin{abstract}
We show that the minimal supergravity extension of the standard model
automatically contains topologically stable electroweak strings if the
hidden sector is invariant under the exact R-symmetry. These defects
appear in the form of the semiglobal R-strings, which necessarily
carry $Z$-flux inside their core. This result is independent from the
particular structure of the hidden sector.

Discussed strings differ fundamentally from the embedded $Z$-strings.
If R-symmetry is explicitly broken (e.g. due to an anomaly), the decay
of the semiglobal strings may have important implications for the
baryogenesis.
\end{abstract}
\newpage

\dspace

\section*{1. Introduction}

According to the present understanding of cosmology, spontaneous breaking
of some gauge symmetry $G_l$ (down to its subgroup $H_l$) may lead to
the formation of the topologically stable vacuum defects. As  is well
known, if the first homotopy group of the vacuum manifold is nontrivial
$\pi_1(G_l/H_l)\neq 1$, these  stable defects are strings. Strings carry a
topologically conserved gauge flux in their core. Thus the above condition
is usually considered as a necessary condition for the existence of the
topologically stable flux inside the defect.

At present the only known example of the spontaneously broken gauge
symmetry in nature is $SU(2)\otimes U(1)$ electroweak group. Therefore,
it is very important to know whether there might exist topologically stable
defects that carry electroweak flux. Since $\pi_1[SU(2)\otimes
U(1)/U(1)_{EM}]$
is trivial, one would normally expect that there can not be
any topologically conserved electroweak flux unless $SU(2)\otimes U(1)$ is
 embedded in some larger local symmetry group (e.g. in grand unification
symmetry). However, we have shown recently that
this assumption is false [1]; topologically stable electroweak flux can easily
exist in the theory provided the global symmetry group $G_{gl}$ is larger
 by at least the spontaneously broken $U(1)_{gl}$ factor under which
electroweak Higgs doublet transforms nontrivialy. Thus, it turns out that
the condition $\pi_1(G_l/H_l) = 1$ does not contradict necessarily
the topological stability of the flux provided the actual exact global
symmetry $G_{gl}$ (which covers $G_l$) broken down to $H_{gl}$ (covering
$H_l$) satisfies $\pi_1(G_{gl}/H_{gl}) = Z$.

 As  was shown in [1], the minimal extension of the standard model in
which topologically stable electroweak strings are presented is an
$SU(2)\otimes U(1)$ gauge theory with  two Higgs doublets and an
$U(1)_{gl}$ global symmetry under which doublets carry different charges.
These strings should not be confused with the embedded $Z$-strings [2]
or/and the semilocal ones [3]. Embedded string solutions are always
topologically unstable (but can be stable under small perturbations in
some range of parameters). It is impossible to obtain our solution from
the embedded (or semilocal) string by continuous change of the parameters
(preserving $SU(2)\otimes U(1)$-symmetry). The reason is simple,
by definition  embedded (or semilocal) strings are just usual
Nielsen-Olesen [4] type vorteces, but embedded into a larger local
(or global) $SU(2)\otimes U(1)$ symmetry. Therefore in the
case of embedded $Z$-strings all vacuum expectation values (VEVs),
independently of the number of Higgs doublets, wind by
$one$ $and$ $the$ $same$ $U(1)_Z$-gauge transformation. In contrast, the
crucial point for the existence of the topologically stable Z-strings
is that the phases of the two doublets should wind $differently$
(since the transformation around the string is never a pure gauge)
 and this guarantees the topological stability of the $Z$-flux.
To avoid possible confusion we will call our strings
``semiglobal'' in order to indicate the role played by the global symmetry.

Of course, the $U(1)_{gl}$ symmetry, if it is exact, has to be broken at
some very high scale (exceeding $10^9 GeV$ or so [5]) by some
$SU(2)\otimes
U(1)$-singlet VEV, otherwise the resulting Goldstone bosons (which
inevitably couple to matter fermions) can not obey astrophysical
constraints. Thus, the minimal realistic Higgs sector allowing
topologically stable electroweak strings includes two doublets and at least
one gauge singlet with a large VEV. Naturally, we would like to know
which are the physically important extensions of the minimal electroweak
model that allow stable strings.

 We find it rather interesting that
conventional $N=1$ supergravity extensions
of the standard model [6] automatically fit in this class of theories.
Due to supersymmetry (SUSY), they contain at least one pair of
Higgs doublets and a number (at least one) of $SU(2)\otimes U(1)$-singlets
with a large VEVs in the ``hidden''  sector (which is responsible for SUSY
breaking). By definition, hidden sector Higgs fields are allowed to
have only gravitational strength interactions (suppressed by powers of
Planck mass $M_{pl}$) with an ``observable'' sector and thus should be
singlets under ``observable'' gauge symmetries.

 In the present paper we  study the existence of the semiglobal
electroweak strings in the locally supersymmetric standard model and point
out some of their possible cosmological consequences. It turns out that in
the minimal case (which does not assume any extra gauge singlets in the
observable sector) the sufficient condition for the existence of these
defects is an exact (or approximate) $R$-symmetry in the hidden sector.

\dspace

\section*{2. Topologically stable flux}

In this section we will briefly recall the mechanism of Ref. [1] leading to
the topological stability of the Z-flux and indicate more explicitly the
difference of this solution with respect to embedded or semilocal strings.
Consider an $SU(2)\otimes U(1)$-theory with  two Higgs doublets
$H$ and $\bar{H}$ with the opposite hypercharges and an extra global symmetry
$U(1)_{gl}$ under  which their charges are $not$ opposite (we use the usual
supersymmetric convention of a doublet and an antidoublet).
 As we said before, $U(1)_{gl}$ has to be broken at
some high
scale  $M_{gl}$ with a VEV of the gauge singlet scalar $S$. The relative
$U(1)_{gl}$-charges of the fields $H$, $\bar{H}$ and $S$ are fixed from
the  explicitly phase dependent couplings in the scalar potential. In the
renormalizable theory there are two alternative terms $SH\bar H$ or
$S^2H\bar H$ (+ h.c.). Let us for definiteness choose the trilinear
one, which in terms of the $U(1)_{EM}$-invariant VEVs  can be written in
the form

\begin{equation}
V_{phase}=\mu v\bar{v}s\cos{(\theta_S-\theta)}
\end{equation}

\noindent

where $\langle H_0\rangle=ve^{i\theta_{H}}$,
$\langle \bar{H}_0\rangle=\bar{v}e^{-i\theta_{\bar{H}}}$ and
$\langle S\rangle=se^{i\theta_{S}}$ are the VEVs of the electrically  neutral
fields and
$\theta=\theta_{\bar{H}}-\theta_{H}$.Furthermore, $\mu$ is a mass parameter
which we will
take to be real and negative. Note, that in order to have a correct value for
the
weak scale ($m_W$), $\mu$ has to be of order $m_W^2/M_{gl}$. This means
that for $M_{gl}$  large, $\mu$ should be very small. However, this choice is
``technically'' natural, since the limit $\mu=0$ enlarges the symmetry by
an extra global $U(1)$.

The VEV
 of the $S$ field breaks $U(1)_{gl}$ and forms the string. The
phase $\theta_S$ winds by
$2\pi n$ (n is an integer) around the string and so does $\theta$, since
$\theta_S$
and $\theta$ are locally correlated through the term (1), which requires
$\theta_S=\theta$. Therefore we have a topological constraint

\begin{equation}
\oint\partial_{\mu}\theta dx^{\mu}=2\pi n
\end{equation}

\noindent

where the integral is taken along the path that encloses the string at
infinity.
The key point is that for the VEVs to be single valued, each of the
phases $\theta_{H}$ and $\theta_{\bar{H}}$ has to wind by $2\pi$-integer.
 So we have

\begin{equation}
\oint\partial_{\mu}\theta_i dx^{\mu}=2\pi n_i
\end{equation}

\noindent

with $n_i$ $(i=H,\bar{H})$ integers that must satisfy a topologically
invariant  condition

\begin{equation}
n_{\bar{H}}-n_{H}=n
\end{equation}

\noindent

Let us take n=1 (corresponding to the minimal global string). The necessary
existence of the flux can be easily seen from the equation of motion for
the $Z$-boson which at the spatial infinity (where all field
strengths vanish) has the form:

\begin{equation}
Z_{\mu}=\frac{\cos{\theta_W}}{g}
\frac{v^2\partial_{\mu}\theta_{H}+\bar{v}^2\partial_{\mu}\theta_{\bar{H}}}
{v^2+\bar{v}^2}
\end{equation}

\noindent

Now taking the integral around the same path and using the condition
(3) and (4) we immediately obtain:

\begin{equation}
Z-flux=\frac{\cos{\theta_W}}{g}
\frac{(v^2+\bar{v}^2)n_{H}+\bar{v}^2}
{v^2+\bar{v}^2}
\end{equation}

\noindent

Obviously, for the minimal string characterized by
n=1 the right hand side of this equation
can never vanish, provided
$v,\bar{v}$ are nonzero. Thus there is
always the $Z$-flux in the core of
the string. This flux is rather unusual. In contrast with an ordinary gauge
string flux it is not a integer so that it can never compensate completely
a logarithmic divergence of the gradient energy at the infinity. These
strings exhibit properties of both global and local strings; they are
semiglobal. Furthermore, although the  flux by itself has no
topological origin, it is topologically stable since the strings are
topologically stable and there can be no string without the flux.

 It is important to understand the fundamental difference between our
semiglobal $Z$-strings and the embedded or/and semilocal
ones. As we have pointed out earlier the key difference comes from the
fact that for the embedded (or semilocal) defect the VEVs wind by
usual local $U(1)_Z$-transformation. This automatically implies that the
phases of all the Higgs doublets (that carry same $SU(2)\otimes U(1)$-
quantum numbers) should wind in the same way. In the sharp contrast
as we have shown, the very existence of the topologically stable
semiglobal strings is just based on the topological constraint (2)
which implies that phases of doublets must wind differently
($n_{H}-n_{\bar{H}}=integer$). In particular, in the n=1 case one expects that
one
of the doublets does not wind at all, so that $U(1)_Z$ gauge symmetry is
not necessarily restored in the core.

 In the case of the semilocal strings $\pi_1(G_l/H_l)$ is nontrivial and
formally there is a topological flux in the gauge sector. This flux however
is never topologically stable since $\pi_1(G_{gl}/H_{gl})$ is trivial and it
costs a finite energy for the flux to be spread out. In the case of the
semiglobal strings situation is just opposite: there is no topological
flux in the gauge sector, but nevertheless the flux is topologically stable
due to the stability of the Higgs configuration.

\dspace

\section*{3. Supergravity extension}

In the minimal N=1 supergravity extensions of the standard model one
assumes the superpotential of the form

\begin{equation}
f=h(S_{\alpha})+w(Y_i)
\end{equation}

\noindent

where $h$ and $w$ are polynomials in the superfields $S_{\alpha}$ and
$Y_i$ belonging to a hidden and observable (quark, lepton, Higgs)
sectors respectively. It is assumed that there is no coupling
among $S$ and $Y$ fields in the superpotential. Therefore, this
two sectors of the theory communicate only through the gravitational
strength interactions in the potential. The most general
renormalizable form of $w$ compatible with
$SU(2)\otimes U(1)$ is

\begin{equation}
w=\mu
H\bar{H}+g_uHQu^c+g_d\bar{H}Qd^c+g_l\bar{H}Le^c+(HL+QLd^c+LLe^c+u^cd^cd^c)
\end{equation}

\noindent

where $H$,$\bar{H}$ are Higgs doublets and $Q$,$u^c$,$d^c$,$L$,$e^c$
are quark and lepton superfields respectively; $g_{u,d,l}$ are
``Yukawa'' coupling constants (family indices are suppressed) and
$\mu$ is a mass parameter which has to be of order $m_W$ for the
correct electroweak symmetry breaking. The terms in the bracket are
usually forbidden by the matter parity symmetry, since if all present they
lead to too rapid proton decay. We keep them here only to emphasize the
generality of our argument, the reader can choose the desired
combination consistent with phenomenological constraints.

The important point  is that the form of the observable
superpotential automatically respects continuous global R-symmetry
$U(1)_R$ under which
for example the R-charges are
 $R_H=R_{\bar{H}}=R_L=1$, $R_Q=1/3$, $R_{u^c}=R_{d^c}=2/3$ nad
$R_{e^c}=0$ (R-charge of the superpotential is normalized as usual to be
$2$). Note that under this R-charge assignment the form of $w$ given by
(7) becomes most general to all orders (including all possible
nonrenormalizable operators).

 We do not want to debate here a currently
popular dilemma whether all global symmetries are necessarily broken by
Planck scale induced operators.  We simply consider both possibilities.
So let us assume for a moment that exact R-symmetry is respected to
all orders in $M_{pl}$. Therefore, for the superpotential (6), the
R-transformation will be a valid symmetry if it is respected by the
hidden sector. If this is the case, R-symmetry will be inevitably broken at
some high scale $M_R$ (not below the scale $M_S$ at which SUSY gets
broken). Breaking occurs due to nonzero VEV of the superpotential
$\langle h\rangle$ (which can never vanish in the nonsupersymmetric minimum
with zero
 cosmological constant) and
due to the VEVs of scalars carrying nonzero R-charges. Spontaneous breaking
of the R-symmetry forms  global R-strings. Effects of the above
breaking on the observable sector can be viewed from the effective
low energy potential of the observable fields which has the following
well known form [7]

\begin{equation}
V=\sum_i|\frac{\partial W}{\partial Y_i}|^2+m_g^*AW^{(3)}+m_g^*BW^{(2)}+h.c.+
\sum_i|m_g|^2|Y_i|^2+(D-terms)
\end{equation}

\noindent

Here $W(Y)=w(Y)\exp(2^{1/2}|S_{\alpha}|^2/M_{Pl})$
is a
redefined low energy superpotential and $W^{(2)}$ and $W^{(3)}$ are its
bilinear and trilinear (in $Y_i$) pieces respectively. $A$ and
$B$ are  numbers related to the details of the hidden sector.

 The message about the R-symmetry breaking from the hidden sector is
carried by the complex parameter $m_g$ whose absolute value is the gravitino
 mass.
It so happens that $m_g$ sets the SUSY breaking scale in the visible
world. Explicit form of $m_g$ is

\begin{equation}
m_g=8{\pi}\langle h\rangle M_{Pl}^{-2} \exp{(2^{1/2}|S_{\alpha}|^2/M_{Pl}^2)}
\end{equation}

\noindent

For us the important thing about $m_g$ (and $\langle h\rangle$)
is that its phase $\theta_h$ winds by $2\pi n$ (with $n=integer$) around
the R-string. This is clear, since $\langle h\rangle$ is polynomial in
condensates
 $\langle S_{\alpha}\rangle$ which must be single valued.
 So the phase dependent coupling in (9) has the following form
(for definiteness $B$ is taken negative):

\begin{equation}
V_{phase}=-|Bm_g\mu H_o\bar{H}_o|\cos{(\theta_h-\theta)}
\end{equation}

\noindent

As before $\theta=\theta_H-\theta_{\bar{H}}$,and since
 $\Delta\theta_h=2\pi n$ around the string, arguments
of Sec.2 are automatically valid. Thus, we conclude that topologically stable
$Z$-flux gets trapped inside the R-string as soon as doublets $H$ and
$\bar{H}$ pick up  nonzero VEVs. This result is independent from the
detailed structure of the hidden sector and the particular mechanism of the
R-symmetry breaking.

\dspace

\section*{4. Discussion on R-symmetry breaking}

In general, the models with spontaneously broken exact  R-symmetry
in the hidden sector may face following difficulties:
(1) Cancellation of the cosmological constant and/or (2) high scale of R-
symmetry breaking which may be cosmologically problematic if $U(1)_R$ is
anomalous.

(1) The first problem is related to the absence of the additive constant in
the  superpotential (unless it is  dynamically generated) which makes it
difficult to adjust the cosmological term to zero. However, the
cancellation of this term can in principle be achieved by means of adjusting
the Kahler potential. An alternative possibility
is to include some strongly coupled gauge interaction that breaks R-symmetry
dynamically (as it is a case in the models with gaugino condensation [8]).

(2) The second difficulty may result from the high scale of R-symmetry
breaking.
Typically the scalar VEVs in the hidden sector are of order $M_{Pl}$ and
can induce $U(1)_R$-breaking at a very high scale. Large $M_R$ can be
cosmologically problematic if R-symmetry is anomalous. This is the case
for instance  in our example of the minimal SUSY standard model in
which R-transformation acts on the quarks as an anomalous Peccei-Quinn
symmetry $U(1)_{PQ}$ [9] and therefore should be broken at a scale
$10^{10}-10^{12} GeV$ [5]. As was stressed in the literature (e.g. see [5]),
this value fits precisely the SUSY breaking scale.  Unfortunately, to use
R-symmetry for the solution of the strong CP-problem (without imposing
additional $U(1)_{PQ}$-symmetries) seems to be problematic, since
$M_R\sim M_S$ leads to the problem of nonzero cosmological
constant. This difficulty comes from the general constraint [10] which says
that the survival of any exact continuous R-symmetry to scales below
$M_R\sim M_S^{2/3}M_{Pl}^{1/3}$ is incompatible with a zero cosmological
constant. Namely, if a continuous R-symmetry
is broken by the nonzero VEV of the superpotential $h$, then the the
cancellation of the cosmological term implies that at the minimum

\begin{equation}
\langle h\rangle \sim M_{Pl}\langle F_S\rangle=M_{Pl}M_S^2
\end{equation}

\noindent

where $\langle F_S\rangle$ is a VEV of the F-term that breaks supersymmetry.
Thus it
looks likely that in theories with a spontaneously broken R-symmetry in the
hidden sector either
 R-symmetry should be exact (nonanomalous) or if it is
approximate, should be explicitly broken by a sufficiently large amount.

 If R-symmetry is anomalous (acts as PQ-symmetry), it should be
explicitly
broken also by some other interaction which gives large enough mass to
the would be R-axion and makes it compatible with cosmology. This explicit
breaking can be induced for example by higher dimensional Planck scale
operators (if we assume that supergravity does not respects R-symmetry)
or/and by some additional metacolor sector  which breaks $U(1)_R$ through
the anomaly. Resulting semiglobal R-strings are no more topologically
stable below the scale of the metacolor phase transition $M_{mc}$ and form
boundaries of  domain walls. The isolated string  bounding
infinite planar  wall is stable for all practical purposes, since the
probability of the hole formation is exponentially suppressed by the ratio
$M_R/m_a$  where $m_a$ is a mass of would be R-axion.
These structures will then, in the ususal manner, decay through the
collapse [11].

\dspace

\section*{5. Implications for Baryogenesis}

Recently, there were some speculations about the possible role of the
unstable vacuum defects in the baryogenesis. In particular in Ref. [12]
it was argued that such a role  can be played by the collapsing loops of the
embedded $Z$-strings. The crucial point however is that to be relevant for
the baryogenesis, embedded strings should be at least quasistable in order
to survive to scales below the electroweak phase transition.
Unfortunately, this is not the case in the minimal standard model with a
single Higgs doublet [13]. The authors of Ref.[12]  assumed
that embedded $Z$-strings might be stable in some realistic extensions
of the minimal scheme, in particular in the two Higgs doublet model, which in
the light of the very  recent analysis [14] is doubtful.
 On the other hand, as we have shown, R-strings in the  SUSY standard
model are topologically stable electroweak $Z$-strings. But, as we said, if
R-symmetry  is anomalous it should be explicitly broken
by sufficiently large amount either by gravity or by some strongly coupled
metacolor sector. In such a case resulting $Z$-strings are no more
topologically stable below the scale $M_{mc}$ where metainstanton
(gravitational or some other) effects become important. But instability is
just what one needs for the baryogenesis if the scale $M_{mc}$ is
sufficiently low so that strings can survive below $m_W$. Below this scale
(while still stable under tunneling) the string network will tend to
decay through the collapse and may produce a certain  baryon to entropy
ratio. The quantitative analysis of this process depends on the detailed
mechanism of R-symmetry breaking and is beyond the scope of the present
paper.

\section*{\bf Acknowledgement}

  We acknowledge useful discussions with Charan Aulakh on the semiglobal
strings in supersymmetry and Riccardo Barbieri on the issue of R-symmetry
breaking.

\end{document}